\documentclass[10pt,conference,letterpaper]{IEEEtran}

\usepackage{amssymb}
\usepackage{amsfonts}
\usepackage[pdftex]{graphicx}
\usepackage{subfigure}
\usepackage{array}
\usepackage{amsthm}
\usepackage[hyphens]{url}
\usepackage{stfloats}
\usepackage{algorithm}
\usepackage{algorithmic}
\usepackage{cite}
\usepackage{balance}
\usepackage{enumerate}
\usepackage{enumitem}

\ifCLASSINFOpdf
\else
\fi
\begin{document}
%
\title{\huge Trailing the Snail: SDN Controller Security Evolution}
%
%
%

\author{\IEEEauthorblockN{Sandra Scott-Hayward}
\IEEEauthorblockA{Centre for Secure Information Technologies (CSIT), Queen's University Belfast, Belfast, BT3 9DT, N. Ireland\\Email: s.scott-hayward@qub.ac.uk}}

\maketitle

\begin{abstract}
The first OpenFlow Software-Defined Network (SDN) Controller, NOX, was developed by Nicira Networks and donated to the research community in 2008. Almost 10 years later, there are at least 29 open-source SDN Controllers and many more proprietary solutions. Two of the open-source SDN controllers stand out in terms of broad deployment and strong contributor base; Open Network Operating System (ONOS) and OpenDaylight (ODL). Both have been deployed in live networks. However, despite increasing adoption of SDN, the security of the SDN control plane has developed at a snail's pace. In this paper, the evolution of ONOS and ODL security is discussed. The reflection of this on secure SDN Controller design is analyzed.
\end{abstract}

%
\section{Introduction}\label{intro}

The first OpenFlow SDN controller was NOX \cite{RefWorks:481}, which was developed by Nicira Networks and donated to the research community in 2008. NOX is a basic platform for building network control applications. With the availability of OpenFlow (OF) and NOX, research and development of SDN-based architectures, network elements and controllers commenced. 

Almost 10 years later, there are at least 29 open-source SDN Controllers \cite{sdnwiki} and many more proprietary solutions. This paper presents a detailed description of the evolution and security status of two of the leading open-source Software-Defined Network (SDN) controllers; Open Network Operating System (ONOS) and OpenDaylight (ODL). The controllers are assessed against a series of security features. ONOS and ODL are relevant indicators of the state of secure SDN controller design based on their large contributor base and broad deployment.

In April 2013, the ODL project \cite{opendaylight} was founded by a group of member companies and hosted by The Linux Foundation. The goal of the project is to accelerate the adoption of SDN and create a solid foundation for Network Function Virtualization (NFV). ODL is to be a common open-source framework and platform for SDN. The first release was in February 2014.

During 2014, the Open Networking Lab (ON.Lab) worked with a number of industry partners and in December 2014 released the first code for ONOS \cite{RefWorks:710}. ONOS was launched as a SDN network operating system for service provider networks with a focus on high availability, scalability and performance. ONOS joined the Linux Foundation in October 2015.

\begin{figure}[htbp]
\centering
    \includegraphics[width=\columnwidth]{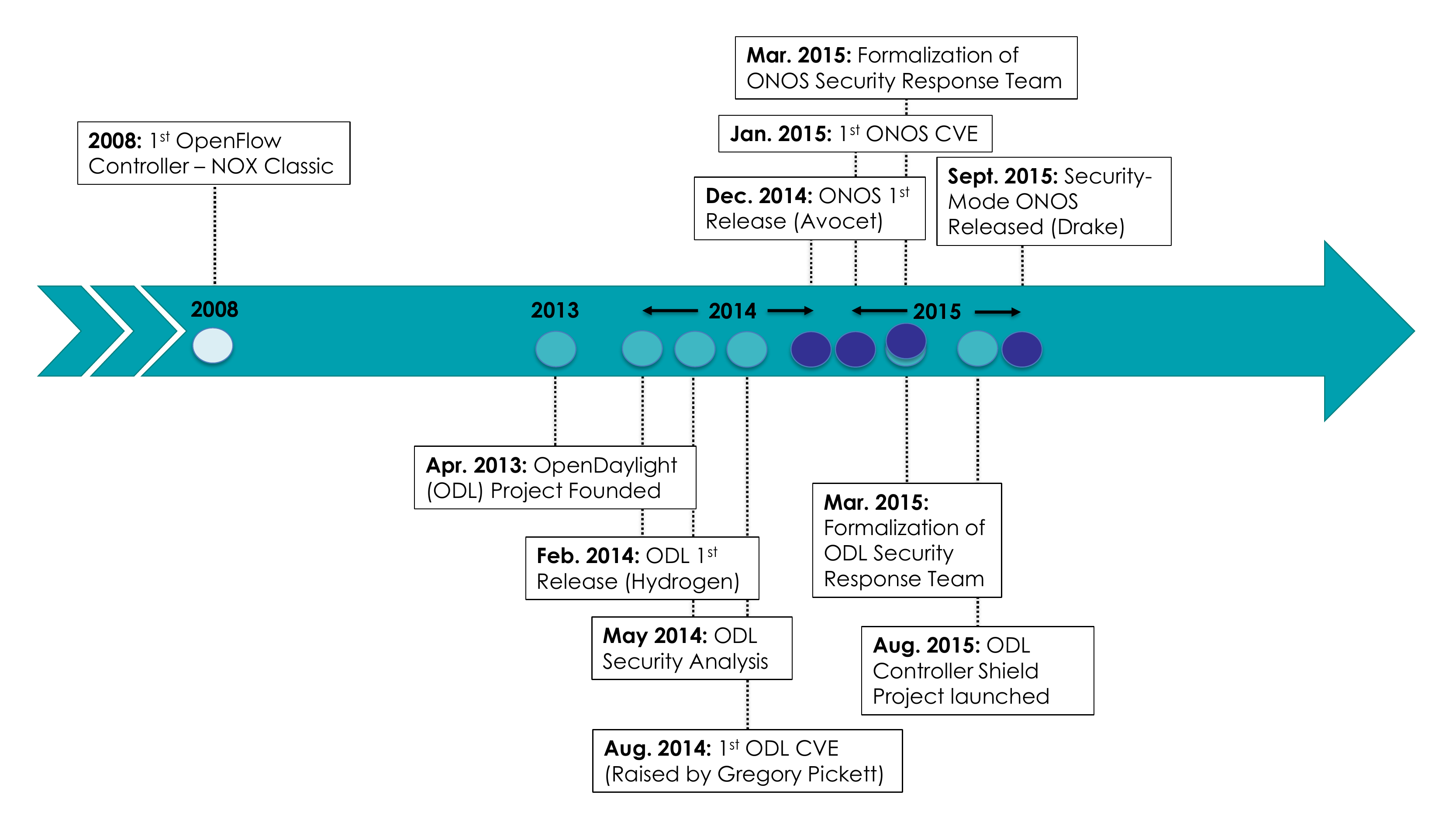}
    \caption{ODL and ONOS security-related events}\label{fig:timeline} 
\end{figure}

In terms of the security of ODL and ONOS, arguably ONOS took a lead from the perspective of resilience with the focus on high availability (controller clustering) from the initial code release. However, during 2014 and 2015, a series of security-related events raised security awareness in both ODL and ONOS. These events are highlighted in Fig. \ref{fig:timeline} and will be discussed in Section \ref{evolution}. 

This paper is organized as follows: Section \ref{background} introduces related work with respect to the features of a secure, robust, and resilient SDN controller. The evolution of secure design of ONOS and ODL is presented in Section \ref{evolution} and in Section \ref{discuss}, the SDN controller security status is analyzed and discussed. Section \ref{conc} concludes the paper. 

\vspace{-2mm}
\vspace{-2mm}

\section{Related Work}\label{background}

A set of 11 features for a secure, robust, and resilient SDN controller were defined in \cite{scott2015design} and are summarized here.

\noindent\textbf{Secure controller design:}
\begin{enumerate}
	\item Control process (application) isolation \textendash{} the ability to separate the application processes running at the controller in order to limit the capability of individual applications to influence control of the network.
	\item	Implementation of policy conflict resolution \textendash{} the ability to resolve incompatible flow rules from multiple apps.
	\item	Multiple controller instances (resilience) \textendash{} to overcome the single, centralized controller as single point-of-failure.
	\item	Multiple application instances (resilience) \textendash{} to handle coordination of network state across multiple app instances.
	\item	Secure storage \textendash{} to protect critical network-related data.
\end{enumerate}
\noindent\textbf{Secure controller interfaces:}
\begin{enumerate}[resume]
	\item Secure control layer communication \textendash{} to protect control communications to/from the data and application planes.
	\item	GUI/REST API security \textendash{} to protect additional interfaces to the controller such as web.
\end{enumerate}
\noindent\textbf{Controller security services:}
\begin{enumerate}[resume]
	\item IDS/IPS integration \textendash{} support for integration of network security applications to the controller.
	\item	Authentication and Authorization \textendash{} support for access control for users, applications, and resources.
	\item	Resource monitoring \textendash{} the ability to protect against any single application consuming excessive resources.
	\item	Logging/Security audit service \textendash{} the collection of appropriate log information to enable security audit.
\end{enumerate}

In the same publication, this author presented a summary of academic work related to the security of SDN/OF controllers. To summarize, in \cite{RefWorks:711}, feature 6 was considered with assessment of controller response to malformed OF messages. Several features were considered in the security-enhanced controller designs of \cite{rosemary, SEFloodlightPaper}. ODL and ONOS were relatively new and there is no reference to security of ONOS in \cite{RefWorks:711,rosemary, SEFloodlightPaper}, while \cite{rosemary} refers to security vulnerabilities in ODL. Since 2015, a number of works have addressed security vulnerabilities/threats to ONOS and ODL controllers and these are discussed here.

Focusing on feature 1, in \cite{noh2015vulnerabilities} a state-based permission system is introduced to examine the permission set of each application and prevent it from executing without permission. The vulnerability of ODL v1.0 to Denial-of-Service (DoS), system shell execution, and data plane poisoning (partially) is demonstrated.

In \cite{adenuga2016security}, the authors provide a limited review on the security threats to SDN controllers focusing on the north/southbound interfaces (feature 6). There is no reference made to the challenge of third-party applications. A few experiments using ONOS v1.7.0 are described illustrating the ability to perform a Man-in-the-Middle (MITM) attack on an unencrypted channel and to perform a DoS attack on the channel. No additional attacks or solutions are described in this report as compared with previous analyses. 

Bidaj uses fuzzing to test network protocols and detect vulnerabilities in \cite{bidaj2016security}. This concerns features 6 and 10. The study considers ODL Lithium-SR3 and Beryllium \cite{opendaylight} and ONOS v1.5. In terms of vulnerabilities, the predominant issue identified through this testing is DoS due to excessive resource consumption; memory or CPU. All five identified vulnerabilities affect ODL. Fuzz testing with manipulated control packets also produced exceptions indicating poor protocol implementation but not raising security issues. Six bugs affected ODL v3.3 and v4.0 while three bugs affected ONOS v1.5. Although the analysis is limited to the known DoS issue, one novel contribution is a risk assessment of the threats based on a scale derived from the ODL and ONOS security advisories. 

In \cite{arbettu2016security}, the authors present a security analysis of ODL, ONOS, Rosemary and Ryu SDN controllers based on the Microsoft STRIDE analysis. They find that ODL is the most secure of these controllers. However, the assessment credits ODL with DoS mitigation, which is not the case, as will be outlined in Section \ref{evolution}. On balance, ODL and ONOS may be closer in security level. The recommendations in \cite{arbettu2016security} reflect those previously identified in \cite{scott2015design}.

The focus of the related work on ODL/ONOS security has been on a subset of the identified features (1,6 and 10). This is also captured in an Appendix to the Open Networking Foundation (ONF) technical recommendation ``Security Foundation Requirements for SDN Controllers'' \cite{secwg-doc1} in which the compliance of a range of open-source controllers against the security requirements is summarized. In the next section, the ODL/ONOS security projects and applications developed by the community are discussed.

\section{Evolution of Secure SDN Controller Design}\label{evolution}

\subsection{Open Network Operating System (ONOS)}\label{onossection}

\subsubsection{Security Support}
As identified in Section \ref{intro}, ONOS was released after ODL and perhaps benefited from some lessons learned in the deployments of early ODL releases. From the outset, ONOS delivered a distributed architecture.  

In January 2015, just following release of ONOS v1.0, ODL was in the SDN media due to a vulnerability that took 4.5 months to be acknowledged and rectified by the developers (see Section \ref{odlsection} for further detail). As a consequence, the first ONOS CVE (Common Vulnerabilities and Exposures) raised in January 2015 was immediately patched. The ONOS security response team was also created in early 2015 and the security response system is described in the project wiki \cite{onoswiki-sec}. Just two CVEs are listed under security advisories for 2015 with a further 4 added in 2017. However, an analysis of the ONOS JIRA for bug tracking identifies a volume of security-related bugs. 

\subsubsection{Security-specific Projects}\label{onosproj} 

There is a separation in ONOS between projects and applications. As such, the two are presented here separately.

\paragraph{\textbf{Security-Mode ONOS (SM-ONOS) \cite{sm-onos}}}
This project adds two main features to ONOS; application authentication and role-based/permission-based access control. SM-ONOS has been developed by KAIST. 

SM-ONOS requires ONOS operators to review and accept a security policy before activating each application. It uses a permission-based approach where an application cannot access a resource unless explicitly granted permission to do so. Application authentication is enabled by the AppID. In order to enable the application to access ONOS services, a set of required permissions must be defined for the application. These permissions relate to the services and APIs that the application will require to interact with ONOS. For example a permission of type APP\_READ provides permission to read various information about installed applications. The permissions required for the application are defined in the application policy file, \textit{app.xml}. Once the application is installed on ONOS with the relevant permissions, it must be reviewed and the requested permissions accepted before the application can be activated. Otherwise activation will raise a policy violation. 

One of the main issues with the SM-ONOS design is the lack of fine-grained access control, effectively granting all access or no access to certain controller features/functions by the application based on two role types - user or admin. A further consideration is that in order to effectively use SM-ONOS, application policy files should be created for all applications in use. The SM-ONOS developers proposed that policy files could be generated for applications using a static analysis approach. This has also been detailed as a feature of ONOS v1.5 (see Table \ref{onossecevolution}). However, to the best of the author's knowledge based on communication with the developers, this work has not and will not be progressed. 

Following release of ONOS v1.6, a bug was identified relating to SM-ONOS interaction with the consistent data store in the ONOS core. Until a patch was issued, SM-ONOS was incompatible with ONOS v1.6. As a result, unit tests have been added from v1.9. However, while SM-ONOS remains outside of the core ONOS build, the project remains unused.

\paragraph{\textbf{Access Control based on DHCP \cite{dhcpaccess}}}
This project was proposed in July 2016 and is included here for completeness. However, there is no application code available or evidence of development. The objective is to control customer port access based on dynamic host configuration protocol (DHCP) snooping. Only DHCP traffic would be allowed for each customer port in ``restricted'' state. Once the DHCP server grants access via DHCP ACK, the customer port would be switched to ``granted'' state with customer traffic allowed. Then, upon DHCP NAK or end of lease, the port/customer is switched back to restricted state. 

\subsubsection{Security Applications}\label{onosapp} 

\paragraph{\textbf{ACL}}
ACL is a built-in ONOS application to build access control lists. The access control list consists of 5-tuple rules to allow or deny IP traffic. The application is integrated with the distributed store to link ACL rules to relevant devices. Various services are provided to map deny rules to allow rules, to identify new ACL rule priority etc. The ACL application was added to ONOS in July 2015 and released with Drake in Sept. 2015. There has been some activity in 2016 with the development of a test tool for the application. However, some bugs remain unresolved \cite{acl-jira}. 

\paragraph{\textbf{AAA}}
In the most recent documentation, dated 11 March 2017, AAA is provided as an app within OpenCORD \cite{aaa_cord}, which is a reference implementation of CORD (Central Office Re-architected as a Datacenter). In previous ONOS versions, the AAA app was part of the core system, \cite{aaa_onos_tutorial}. The intention of the AAA app is that it is the first element that checks incoming traffic, leveraging the RADIUS server to either block unauthorized traffic or open flows on the switch to enable authenticated and authorized traffic. At the time of writing, the AAA app only performs authentication. Based on conversations in the ONOS mailing lists, this app is a placeholder for future community development. 

The estimated maturity of each ONOS security project and application described in this section is noted in Table \ref{onos-project-maturity}. The maturity is based on assessment of the functionality provided and ease of implementation of the solution.

\vspace{-2mm}
\begin{table}[htpb]
\scriptsize
\renewcommand{\arraystretch}{1.2}
\caption{Estimated Maturity of ONOS Security Projects/Applications}
\begin{center}
\begin{tabular}{| l || l | l | l |}
\hline\hline\noalign{\smallskip}
\textbf{Project/} & \textbf{Proposal} & \textbf{Estimated } & \textbf{Comment}\\
\textbf{Application} & \textbf{Date} & \textbf{Maturity} & \\
\noalign{\smallskip}
\hline\hline
\noalign{\smallskip}
SM-ONOS & Jan. 2015 & Medium & Functional but limited practical \\
& & & use as not part of core ONOS \\
\hline
AC - DHCP & Jul. 2016 & N/A & Project description only\\
\hline
\hline
ACL & Jul. 2015 & Low & Application developed but \\
& & & implementation issues logged \\
\hline
AAA & Sept. 2015 & Low & Authentication only; \\
& & & Not part of core ONOS \\
\noalign{\smallskip}
\hline\hline
\end{tabular}
\end{center}
\label{onos-project-maturity}
\end{table}
\vspace{-2mm}

\subsubsection{Security-focused design}

Since the first version of ONOS, there have been a number of security improvements. As highlighted in Section \ref{intro}, although resilience in terms of high availability was a design focus from the beginning, in the first version of ONOS, AAA was largely missing. There has been a progressive improvement in the ONOS security features with each new software version. For example, GUI/REST API/CLI security settings and TLS/SSL support are now included. 

\begin{table*}[htpb]
\scriptsize
\renewcommand{\arraystretch}{1.2}
\caption{Evolution of Security Features in ONOS Design}
\begin{center}
\begin{tabular}{| l || l | l |}
\hline\hline\noalign{\smallskip}
\textbf{Version} & \textbf{Release Date} & \textbf{Security Features} \\
\noalign{\smallskip}
\hline\hline
\noalign{\smallskip}
Avocet (v1.0) & December 2014 & High Availability \\
\hline
Blackbird (v1.1) & February 2015 & \\
\hline
Cardinal (v1.2) & May 2015 & \\
\hline
Drake (v1.3) & September 2015 & GUI and CLI require username and password login; REST interfaces require username and password; TLS support\\
& & for inter-node communication; Configurable HTTPS for GUI and REST API; Security-Mode ONOS for application security \\
\hline
Emu (v1.4) & December 2015 & \\
\hline
Falcon (v1.5) & March 2016 & \textit{[Automatic application security policy extraction using static analysis techniques (KAIST)]}; SecurityGroup feature of OpenStack\\
\hline
Goldeneye (v1.6) & May 2016 & \\
\hline
Hummingbird (v1.7) & September 2016 & \textit{[New subsystem for anomaly detection (ATHENA \cite{lee2017athena}) (SRI)]}; Rate limit on port via NetConf (GEANT)\\
\hline
Ibis (v1.8) & November 2016 & \\
\hline
Junco (v1.9) & February 2017 &  Implemented unit test for Security-Mode ONOS, Integrated Security (DELTA) tests into OnosSystemTest \\
\hline
Kingfisher (v1.10) & June 2017 & Added support of security group to openstack/networking-onos and SONA (Simplified Overlay Network Architecture) \cite{onos-sona}\\
\noalign{\smallskip}
\hline\hline
\end{tabular}
\end{center}
\label{onossecevolution}
\end{table*}

However, as described in the previous sections, features that would offer protection against some of the threats/attacks described in \cite{rosemary,noh2015vulnerabilities}, are not fundamental to the control design but rather offered as applications. Although authentication mechanisms are now provided, there is no authorization mechanism implemented in ONOS. So, for example, it is not possible to distinguish user rights such as allowing a user to view all applications but disallow the same user to start/stop/remove applications. The evolution of the security features/functionality in ONOS is detailed in Table \ref{onossecevolution}.

\subsection{OpenDaylight (ODL)}\label{odlsection}

\subsubsection{Security Support}

An initial ODL security analysis was performed in May 2014 \cite{odlsecrep}. The report details a series of security features, summarized in Table \ref{odlsecrep}. However, there was little progress in implementing the recommendations. This is particularly evident from the recommendation in the conclusion of the report ``to have a special mechanism to report security bugs/issues to OpenDaylight''. In August 2014, Gregory Pickett, a security researcher at Hellfire Security, discovered a vulnerability in ODL, which he flagged to the developers. It took until January 2015 for a patch to be generated. Following this event, ODL introduced a public security mailing list, a security advisories web page and a security response team.  

In January 2015, David Jorm presented ODL's security vision to be an industry-leading, secure engineering function with security documentation, developer training for security, and automated quality engineering/continuous integration jobs to catch issues and regressions \cite{odlsecpresi}. In July 2015, Jorm presented further updates on ODL secure development processes \cite{odljormJul15} and introduced the Secure Engineering Intern project with objectives of documenting best practices and a threat model, automating checks for known-vulnerable dependencies, and automating static analysis checks. The project appears to have been allocated but no results have been reported.  

Details of ODL security systems and efforts are provided at \cite{odlsecpage}. The ODL Security Advisories web page \cite{odlsecwiki} lists all the security vulnerabilities fixed in ODL. The list is active.
\begin{table*}[htpb]
\scriptsize
\renewcommand{\arraystretch}{1.2}
\caption{Summary of ODL Security Features, May 2014 \cite{odlsecrep}}
\begin{center}
\begin{tabular}{| l || l | l |}
\hline\hline\noalign{\smallskip}
\textbf{Security Feature} & \textbf{Comment} & \textbf{Recommendation} \\
\noalign{\smallskip}
\hline\hline
\noalign{\smallskip}
Application Bundle Security & Bundles provide some level of isolation & Augment with bundle signature/permission verifiers at loadtime, bundle access\\
& & security at runtime; Bundle authentication/authorization should be logged \\
\hline
OSGi Runtime Container Security & Concerns with security footprint of Karaf & Make Karaf security documentation available to ODL developers \\
- ODL Apache Karaf Distribution & & and administrators \\
\hline
ODL Controller Plugins Security & Secure communication access to the controller; & Provide secure access for 8 plugins, DDoS attack protection on \\
& 5/13 plugins use secure versions of protocol & plugin exposed ports, use a common crypto key storage, \\
& & and support pluggable/built-in CA \\
\hline
AAA for External Users & Supports secure access via NB API & Provide role-based access control for external users, user access\\
& & authentication, access protocol authorization, services/resource\\
& & authorization, auditing access/authorization pluggable AAA service\\
\hline
Secure Device/Controller BootStrap & Controller/Device Discovery is manual & Zero-touch bootstrap requirements - automatic device discovery \\
Authentication and Authorization & & and AAA support \\
\hline
Controller Clustering and Security & Clustering comms channel should be secure & Configure Jgroups AUTH and ENCRYPT support for security \\
\noalign{\smallskip}
\hline\hline
\end{tabular}
\end{center}
\label{odlsecrep}
\end{table*}

\subsubsection{Security-specific Projects}

Each of the security-related projects of ODL are discussed here.

\paragraph{\textbf{Controller Shield \cite{CS-project}}}
As highlighted in Figure \ref{fig:timeline}, ODL launched this project in August 2015 just before the 1st code release of SM-ONOS. The objective of the project is to collate security-related information for protection of the controller and expose this information to northbound applications. It is comprised of two elements: 1) Unified Security Plugin (USecPlugin): This plugin registers for Packet-In notifications and calculates packets-per-second (pps) on a flow basis. An alarm is generated if a pps threshold is exceeded and the alarm is exposed to northbound applications. The USecPlugin was proposed for the ODL Beryllium version with extension in the Boron version. 2) East-West interface protection: With this service, the controllers are authenticated before an inter-controller BGP session is initiated. This service was proposed for release with Boron. 

The code for USecPlugin has been developed. However, the east-west interface protection has not been progressed and there is no recent activity.

\paragraph{\textbf{Cardinal - ODL Monitoring as a Service \cite{cardinal-odl}}}
This project is noted here as it offers a fault/health monitoring service, which could be extended or integrated for security services. For example, the provision of CPU/memory usage information for resource monitoring related to network attack. The objective of Cardinal is to enable legacy network management systems to inter-work with ODL. Cardinal consists of an abstraction layer for exposing monitoring, diagnostics and events to northbound monitoring and analytics applications. The adaptor has been developed and supports snmpd, snmptrapd agents with planned enhancements for SNMPv2c trap and info messages, and further versions/agents.

\paragraph{\textbf{AAA \cite{odl-aaa}}}
The AAA project was launched in ODL mid-2014. The identity of users is authenticated, and user access to resources is authorized and recorded. The user authenticates to the controller with a username/password combination and receives an access token to access protected resources on the controller. Access to specific resources is determined by the user role and permissions. ODL leverages the Apache Shiro Java security framework to provide AAA services. Shiro provides four elements: (1) Authentication, (2) Authorization, (3) Cryptography, and (4) Session Management. ODL uses priority ordering to ensure that the AAA function processes incoming requests before any other filtering is performed. It is assumed that the AAA security implementation will not change significantly with future ODL versions, based on the stability of the documentation from prior releases. The change logs indicate that the AAA app is still being developed, with recent bug fixes available.

\paragraph{\textbf{Defense4All \cite{d4a}}}
Defense4All (D4A) was the first security application developed for ODL. It was submitted by RadWare in 2013 as an open-source version of their DefenseFlow product for detecting and mitigating DDoS attacks. D4A was released with Hydrogen and provides a system for detecting attack traffic and redirecting it based on ODL's monitoring and control capabilities. It operates by using OF flow entries to capture flow statistics; coarse flows at the network edge for high volume detection, and fine-grained flows near protected servers. Traffic redirection is also implemented with OF flow entries. D4A provides the attack detection while attack mitigation requires specialized equipment. The mitigation manager in D4A supports the selection and configuration of these external devices. It is noted that the D4A REST API does not check for credentials or define user roles to authorize usage of certain REST APIs. There has been no activity on the Defense4All application since January 2015. 

\paragraph{\textbf{Secure Network Bootstrapping Interface (SNBI) \cite{odl-snbi}}}
SNBI is used for securely bootstrapping the network infrastructure by automating the setup process for required devices and their credentials. The SNBI project was proposed in May 2014. The SNBI process begins with Neighbour Discovery (using IPv6 addressing). Once a neighbour is discovered, the protocol requests the device domain certificate to verify if the device is already bootstrapped. If it is not, the bootstrap of the new device is initiated by the proxy device. 

With SNBI, the registrar is hosted in the SDN Controller and is the trusted entity in the network domain. The SDN controller also contains a plugin for the secure discovery service to communicate with the network devices. Network devices must run the SNBI software package for the secure discovery service. New devices then discover the domain by searching for the SNBI registrar, and present their 802.1AR credentials. Once authorization is confirmed, the new device enrols with the domain and establishes a secure communication channel. This process assumes same domain membership of the device and the SNBI registrar.

The first release of the SNBI solution was with Helium with additional features added in Lithium and Boron. There is no further development plan but the project is maintained.

\paragraph{\textbf{Unified Secure Channel (USC) \cite{usc-project}}}
The objective of this project is to provide a unified secure communication tunnel between the network element and controller to guarantee unified bidirectional secure communication across unsecured public infrastructure. According to the project scope, USC should offer a secure channel with two-way initiation, two-way authentication, and should support multiple protocols in parallel. This project has been contributed by Huawei Enterprise and was released with Lithium. The project is still active with commitment to the Carbon release. 

The estimated maturity of the ODL security projects is noted in Table \ref{odl-project-maturity}. The maturity is based on assessment of the functionality provided. 

\vspace{-2mm}
\begin{table}[htpb]
\scriptsize
\renewcommand{\arraystretch}{1.2}
\caption{Estimated Maturity of ODL Security Projects}
\begin{center}
\begin{tabular}{| l || l | l | l |}
\hline\hline\noalign{\smallskip}
\textbf{Project} & \textbf{Proposal} & \textbf{Estimated } & \textbf{Comment}\\
& \textbf{Date} & \textbf{Maturity} & \\
\noalign{\smallskip}
\hline\hline
\noalign{\smallskip}
Defense4All & Aug. 2013 & Medium & Solution available, not maintained \\
\hline
SNBI & May 2014 & High & Solution developed and maintained \\
\hline
AAA & Jun. 2014 & High & Extended functionality \\
\hline
USC & Dec. 2014 & High & Solution developed, ongoing updates \\
\hline
Controller & Aug. 2015 & Low & Basic functionality, \\
Shield & & & no recent development \\
\hline
Cardinal & Mar. 2016 & High & Solution developed, ongoing updates \\
\noalign{\smallskip}
\hline\hline
\end{tabular}
\end{center}
\label{odl-project-maturity}
\end{table}
\vspace{-2mm}

\subsubsection{Security-focused design}

There is clear progress towards the integration of security features and functionality in ODL. Both the number of security-related projects detailed in the previous section and the attention of the security response team is evidence of this. However, due to the more distributed (and larger) community of developers of ODL, it is hard to determine the extent of adoption or deployment of these security services beyond their development team. Only 1 of 3 project leaders responded to an inquiry regarding the current/planned development activities and the use of the tools by ODL users. The evolution of the security features/functionality in ODL are detailed in Table \ref{odlsecevolution}. 

\begin{table*}[htpb]
\scriptsize
\renewcommand{\arraystretch}{1.2}
\caption{Evolution of Security Features in ODL Design}
\begin{center}
\begin{tabular}{| l || l | l |}
\hline\hline\noalign{\smallskip}
\textbf{Version} & \textbf{Release Date} & \textbf{Security Features} \\
\noalign{\smallskip}
\hline\hline
\noalign{\smallskip}
Hydrogen & February 2014 & Defense4All DDoS attack detection and mitigation tool\\
\hline
Helium  & September 2014 & \\
\hline
Lithium & June 2015 & \textbf{New features for security and automation:} \\
& & \textit{Unified Secure Channel} eases secure communication between ODL and widely distributed networking equipment; \\
& & \textit{Time Series Data Repository (TSDR)} enables collection and analysis of large amounts of network activity;\\
& & \textit{Device Identification and Driver Management (DIDM)} provides end users the ability to discover, manage and automate a wide range\\
& & of existing hardware in their infrastructure; \\
& & \textit{Persistence} ensures application-specific data is preserved over time or in the event of a catastrophe;\\
& & \textit{Topology Processing Framework} allows for filtered and/or aggregated views of a network, including multi-protocol, underlay/overlay \\
\hline
Beryllium & February 2016 & \textbf{New features for performance and scalability:} \\
& & Stronger analysis and testing of clustering, applications that want to be cluster-aware can choose how to put data across the cluster;\\
& & Fully support OpenStack High Availability and Clustering\\
\hline
Boron & September 2016 & \textit{NetVirt project} enhanced support in OpenStack environments for IPv6, Security Groups (via OpenFlow configuration) and VLANs. \\
& & \textit{Cardinal} project monitors the health of the controller, delivered as a service to existing, deployed network monitoring and analytics tools. \\
& & \textit{Centinel} analytics engine enables end-to-end data collection and machine learning to support performance monitoring and bandwidth\\
& & management across WAN links \cite{centinel-odl}.\\
\hline
Carbon & June 2017 & \textit{NetVirt and Genius projects} integrate to dynamically create and manage tunnels and virtual network functions on demand. \\
\noalign{\smallskip}
\hline\hline
\end{tabular}
\end{center}
\label{odlsecevolution}
\end{table*}

\section{Discussion}\label{discuss}
The security support, projects, applications, and the security-focused design of ONOS and ODL have been detailed in Section \ref{evolution}. The objective of this paper is not to provide a direct comparison of these two open-source controllers from a security perspective, rather to offer an assessment of the state of security of open-source controller design with ONOS and ODL being the most advanced of those deployed.

With regard to the controller security attributes described in Section \ref{background}, there has been an improvement in security features across both controllers since their initial releases, specifically considering AAA functionality (features 9 and 11). However, the implementation is weak in places. For example, the default username and password are easily obtained from the online documentation. In the case where the default credentials are active, there is unlimited access to the controller. If the username/password has been changed, it may still not be too difficult to gain access to the controller. For example, there is no defence against a simple brute-force attack to capture the controller credentials. Unsuccessful attempts to access the REST interface are not logged, there is no limit on the number of unsuccessful attempts that can be made to connect to it nor is there any concept of blacklisting certain connections that continuously attempt to connect to it. In fact, some of the 2017 CVEs highlight just such vulnerabilities \cite{onoswiki-advice, odlsecwiki}.

The lack of progress on elements such as resource monitoring (feature 10) and application isolation (feature 1) is disappointing. These are controller security issues that have been raised repeatedly. This lack of real progress on security is also reflected in the maturity level of the security projects/applications described in Tables \ref{onos-project-maturity} and \ref{odl-project-maturity}. Several of the projects lack development progress. Of the higher maturity projects, it has not been possible to capture statistics regarding the adoption rate. 

As noted in Section \ref{background}, in July 2016, the document ``Security Foundation Requirements for SDN Controllers'' \cite{secwg-doc1} was published by the Open Networking Foundation (ONF) Security Working Group (SecWG). This document provides a detailed breakdown of security requirements derived from a threat analysis of the SDN controller. However, an inquiry to ON.Lab indicated that they were not aware of the document while no response was received from the ODL security list.

The drive in the SDN community (strongly driven by ONF) is towards open-source software rather than standards development. The OSSDN Project Delta \cite{delta} has also emerged from the ONF SecWG. Delta provides a penetration testing framework for SDN, which predominantly explores the impact of the threats in SDN on the controller. Positive feedback has been received since the launch of Delta in 2016. Education of developers on the importance of secure SDN controller design is perhaps better served by the presentation of tools such as Delta. With hands-on tools, the vulnerability of the controller to specific attack types/threats can be demonstrated, which can have a greater impact than a written recommendation. 

However, ownership of the secure design/development must be taken by the core ONOS/ODL teams. The concept of these security projects and applications succeeds to design security features for the controllers. However, it fails when it comes to addressing the fundamental security of the controller architecture and the goal of ``secure by design''.

\vspace{-1mm}
\section{Conclusions}\label{conc}

The security evolution of two leading open-source SDN controllers; ONOS and ODL has been presented in this paper. It is encouraging to see an increasing focus on security within both controller communities through security support and new features introduced with each new software release. Of concern, however, is the lack of integration of security as a core feature of the controller. While there are many project/application proposals for secure controller functionality, there is limited implementation of secure controller design. These controllers must be hardened for deployment in public networks. 

\vspace{-4mm}
\bibliographystyle{IEEEtran}
\bibliography{IEEEabrv,trailsnail_refs}

\end{document}